\documentclass[12pt, a4paper, fleqn]{article}

\usepackage[tbtags]{amsmath}
\usepackage{amssymb}
\usepackage{mathrsfs}
\allowdisplaybreaks[1]

\newcommand{\ui}{\mathrm{i}\;\!}

\newcommand{\cP}{\mathcal{P}}
\newcommand{\cJ}{\mathcal{J}}
\newcommand{\cH}{\mathcal{H}}
\newcommand{\cG}{\mathcal{G}}
\newcommand{\cM}{\mathcal{M}}
\newcommand{\cK}{\mathcal{K}}
\newcommand{\cD}{\mathcal{D}}

\renewcommand{\vec}[1]{\mathbf{#1}}
\newcommand{\nc}[1]{\widehat{#1}}
\newcommand{\ncg}[1]{\widetilde{#1}}

\begin{document}


\title{\textbf{Deformed Schr\"odinger symmetry on noncommutative space}}

\author{
 \textbf{Rabin Banerjee}\\[0.5ex]
 \textit{\normalsize S.~N.~Bose National Centre for Basic Sciences,}\\[-0.2ex]
 \textit{\normalsize JD Block, Sector 3, Salt Lake, Kolkata 700098, India}\\
 {\normalsize E-mail: \texttt{rabin@bose.res.in}}
}

\date{}

\maketitle

\begin{abstract}
{We construct the deformed generators of Schr\"odinger symmetry consistent with noncommutative space. The examples of the free particle and the harmonic oscillator, both of which admit Schr\"odinger symmetry, are discussed in detail. We construct a generalised Galilean algebra where the second central extension exists in all dimensions. This algebra also follows from the Inonu--Wigner contraction of a generalised Poincar\'e algebra in noncommuting space.}
\end{abstract}

\bigskip



\noindent
The introduction of noncommuting relativistic coordinate spacetime,
\begin{equation}\label{101}
\left[\nc{x}^\mu, \nc{x}^\nu\right] =  \ui\theta^{\mu\nu}, \quad \mu, \nu = 0, i,
\end{equation}
for constant $\theta^{\mu\nu}$ implies, among other things, a breakdown of Lorentz invariance. From an algebraic point of view, the Jacobi identity involving the angular momentum operator and the noncommuting coordinates is violated.

Recently it has been found by Wess~\cite{Wess-04} and collaborators~\cite{Dimitrijevic:2004rf, Asch-05, Koch-05} that, by appropriately deforming the classical Poincar\'e generators, consistency with \eqref{101} is achieved while preserving the original Poincar\'e algebra. In other words, a new representation of the Poincar\'e algebra that is compatible with \eqref{101} has been obtained. But the coproduct rules are modified. Also, the modified coproduct rules agree with those found \cite{Chai-05, Matlock:2005zn} from another (quantum-group theoretic) approach based of the application of twist functions~\cite{Oeckl:2000eg}. The extension of these ideas to field theory and possible implications for Noether symmetry are discussed in \cite{Banerjee:2004ev, Gonera:2005hg}. Very recently, the deformed Poincar\'e generators for a Lie algebraic $\theta$ (rather than a constant $\theta$) have also been analysed~\cite{Luki-05}.

In this {paper} we consider the invariance of the Schr\"odinger group \cite{Nied-73:Hage-72:Burd-72} (this contains, in addition to the centrally extended Galilei group, two conformal generators, namely dilatations and special conformal transformations or expansions) on nonrelativistic noncommutative space,
\begin{equation}\label{102}
\left[\nc{x}^i, \nc{x}^j\right] =  \ui\theta^{ij}.
\end{equation}
There are some good reasons for pursuing such an investigation. The question of this invariance is interesting in its own right. Also, the Schr\"odinger group is an entirely consistent group of spacetime transformations which in some respects has a more complex structure than the Poincar\'e group. Finally, the second central extension \cite{Levy-72:Grig-96:Bose-95} of the Galilei group is not well understood and it would be worthwhile to see if this aspect can be illuminated from the deformations.

In this {paper} we consider the free particle and the harmonic oscillator. Although both these models admit Schr\"odinger symmetry, the structure of generators is quite different. The deformed generators compatible with \eqref{102} are computed. These generators satisfy the standard Schr\"odinger algebra. Thus as happens for Poincar\'e symmetry, a new representation of the Sch\"odinger algebra is found that is consistent with \eqref{102}.

We also discuss some applications of the general formalism. The structure of the deformed generators is computed in both the coordinate and momentum representations. In the coordinate picture the deformations are still there but in the momentum picture these cancel out and the functional form of the generators is identical to that obtained for usual (commutative space) theory. Implications of this result are discussed.

We have next considered a more general one-parameter class of deformed generators. It leads to a closure of the algebra which is however more general and hence distinct from the standard Galilean algebra. This generalised algebra is also derived from a contraction of the deformed Poincar\'e algebra recently discussed in \cite{Koch-05}. Moreover, it is found that the boosts do not commute. This (second) central extension is valid in any dimensions. It is a new result since the second central extension is found only in two dimensions \cite{Levy-72:Grig-96:Bose-95, Duva-00, Jack-00, Horvathy:2002vt}. For the special case of two dimensions, we show that our generalised algebra reduces to the standard Galilean algebra, including the second central extension.

This paper is broadly divided into two sections; in the first the formulation is set up while the second is devoted to applications.

\subsection*{1. Deformed Schr\"odinger symmetry}

The $n$-dimensional Schr\"odinger algebra is given by augmenting the Galilean algebra involving rotations ($\mathcal{J}^{ij}$), translations ($\mathcal{P}^i$), Hamiltonian ($\mathcal{H}$) and boosts ($\mathcal{G}^i$) with the algebra of dilatations ($\mathcal{D}$) and expansion or special conformal transformations ($\mathcal{K}$). The relations are
\begin{align}
&\nonumber \left[\mathcal{J}^{ij}, \mathcal{J}^{k\ell}\right] = - \ui\left(\delta^{kj}\mathcal{J}^{i\ell} - \delta^{ki}\mathcal{J}^{j\ell} + \delta^{\ell j}\mathcal{J}^{ki} - \delta^{\ell i}\mathcal{J}^{kj}\right)\\
\nonumber &\left[\mathcal{P}^i, \mathcal{P}^j\right] = 0 & &\left[\mathcal{G}^i, \mathcal{G}^j\right] = 0\\
\nonumber &\left[\mathcal{P}^i, \mathcal{J}^{jk}\right] = - \ui\left(\delta^{ij}\mathcal{P}^{k}-\delta^{ik}\mathcal{P}^{j}\right) & &\left[\mathcal{G}^i, \mathcal{J}^{jk}\right] = - \ui\left(\delta^{ij}\mathcal{G}^{k}-\delta^{ik}\mathcal{G}^{j}\right)\\
\nonumber &\left[\mathcal{P}^i, \mathcal{G}^j\right] = - \ui\delta^{ij}\mathcal{M} & &\\
\nonumber &\left[\mathcal{H}, \mathcal{P}^i\right] = 0 & &\left[\mathcal{H}, \mathcal{G}^i\right] = - \ui\mathcal{P}^i\\
\nonumber &\left[\mathcal{J}^{ij}, \mathcal{H}\right] = 0 & &\\[5pt]
\nonumber &\left[\mathcal{J}^{ij}, \mathcal{D}\right] = 0 & &\left[\mathcal{J}^{ij}, \mathcal{K}\right] = 0\\
\nonumber &\left[\mathcal{D}, \mathcal{G}^i\right] = - \ui\mathcal{G}^i & &\left[\mathcal{K}, \mathcal{G}^i\right] = 0\\
\nonumber &\left[\mathcal{D}, \mathcal{P}^i\right] = \ui\mathcal{P}^i & &\left[\mathcal{K}, \mathcal{P}^i\right] = \ui\mathcal{G}^i\\
\nonumber &\left[\mathcal{H}, \mathcal{D}\right] = - 2\ui\mathcal{H} & &\left[\mathcal{H}, \mathcal{K}\right] = - \ui\mathcal{D}\\
\label{103} &\left[\mathcal{D}, \mathcal{K}\right] = - 2\ui\mathcal{K} & &
\end{align}
where $i, j, \ldots = 1, 2, 3, \ldots n$, and the mass parameter $\mathcal{M}$ commutes with everything.

The standard free-particle representation of this algebra is given by
\begin{align}
\nonumber &\mathcal{J}^{ij} = x^{i}p^{j} - x^{j}p^{i} & &\text{(angular momentum)}\\
\nonumber &\mathcal{H} = \frac{\vec{p}^2}{2m} & &\text{(energy)}\\
\nonumber &\mathcal{D} = \frac{1}{2}\left(p^{i}x^i + x^{i}p^i\right) - \frac{\vec{p}^2}{m}t & &\text{(dilatations)}\\
\nonumber &\mathcal{K} = \frac{1}{2}m\left(x^i-\frac{p^i}{m}t\right)^2 & &\text{(expansions)}\\
\nonumber &\mathcal{G}^i = mx^i-p^i t & &\text{(Galilean boosts)}\\
\nonumber &\mathcal{P}^i = p^i & &\text{(linear momentum)}\\
\label{104} &\mathcal{M} = m & &\text{(mass)}
\end{align}
Using the (commutative-space) canonical brackets,
\begin{gather}
\nonumber \left[x^i, x^j\right] = \left[p^i, p^j\right] = 0,\\
\label{105} \left[x^i, p^j\right] = \ui \delta^{ij},
\end{gather}
the algebra \eqref{103} is easily reproduced from \eqref{104}.

For noncommutative space, Eq.~\eqref{102} has to be augmented by the algebra:
\begin{gather}
\nonumber \left[\nc{p}^i, \nc{p}^j\right] = 0,\\
\label{106} \left[\nc{x}^i, \nc{p}^j\right] = \ui \delta^{ij},
\end{gather}
which conform to the usual form \eqref{105}.

As is known from an analysis of Poincar\'e generators, the angular momentum operator in \eqref{104} is inconsistent with the noncommutative algebra \eqref{102} since the $\cJ$--$\nc{x}$--$\nc{x}$ Jacobi identity is violated if the undeformed transformation law is taken:
\begin{equation}\label{107}
\left[ \cJ^{ij}, \nc{x}^k \right] = \ui \left(\delta^{ik}\nc{x}^j - \delta^{jk}\nc{x}^i\right).
\end{equation}
An appropriate deformation of the the transformation law is necessary to restore the Jacobi identity. It is simple to check that if \cite{Wess-04, Koch-05}
\begin{equation}\label{108}
\left[\nc{\cJ}^{ij}, \nc{x}^k\right] = \ui\left(\delta^{ik}\nc{x}^j - \delta^{jk}\nc{x}^i + \frac{1}{2}\theta^{ik}\nc{p}^j - \frac{1}{2}\theta^{jk}\nc{p}^i + \frac{1}{2}\delta^{ik}\theta^{jm}\nc{p}^m - \frac{1}{2}\delta^{jk}\theta^{im}\nc{p}^m\right)
\end{equation}
then the $\nc{\cJ}$--$\nc{x}$--$\nc{x}$ Jacobi identity is indeed satisfied. A representation for $\nc{\cJ}^{ij}$ that yields the above relation is given by \cite{Wess-04, Koch-05}
\begin{equation}\label{109}
\nc{\cJ}^{ij} = \nc{x}^i\nc{p}^j - \nc{x}^j\nc{p}^i + \frac{1}{2}\theta^{im}\nc{p}^m\nc{p}^j - \frac{1}{2}\theta^{jm}\nc{p}^m\nc{p}^i.
\end{equation}
Note that this deformed operator also satisfies the usual angular momentum algebra:
\begin{equation}\label{110}
\left[\nc{\cJ}^{ij}, \nc{\cJ}^{k\ell}\right] = - \ui\left(\delta^{kj}\nc{\cJ}^{i\ell} - \delta^{ki}\nc{\cJ}^{j\ell} + \delta^{\ell j}\nc{\cJ}^{ki} - \delta^{\ell i}\nc{\cJ}^{kj}\right).
\end{equation}
This is verified by using the basic commutators \eqref{102} and \eqref{106}.

It should be mentioned that a more general transformation law \eqref{108} involving coefficients $\lambda_1$ and $\lambda_2$ is possible \cite{Koch-05} which is also compatible with \eqref{102}. That would lead to a more general structure of the angular momentum operator \eqref{109}. However the choice \eqref{109} is singled out since it alone satisfies the standard closure of the angular momentum algebra as given in \eqref{110}.

We now obtain the deformation of the other generators. The linear momentum $\nc{p}^i$ and Hamiltonian $\vec{\nc{p}}^2/2m$ retain their original forms, basically because the algebra of $\nc{p}^i$ is identical to $p^i$. For boosts ($\cG$) a deformation is necessary. Considering the minimal (i.e. least order in $\theta$) deformation, we obtain the following structure:
\begin{equation}\label{111}
\nc{\cG}^i = m\nc{x}^i - t\nc{p}^i + \lambda_1 m\theta^{ij}\nc{p}^j + \lambda_2 m^3 \theta^{ij}\nc{x}^j,
\end{equation}
where $\lambda_1$ and $\lambda_2$ are, as yet, undetermined coefficients. In order that the standard Galilean algebra involving the rotations, translations and boosts is retained we find $\lambda_1 = 1/2$, $\lambda_2 = 0$. Also, the consistency of \eqref{111} with \eqref{102} is easily established as the relevant Jacobi identity $\nc{\cG}$--$\nc{x}$--$\nc{x}$ is trivially satisfied.

Proceeding in a similar manner deformed dilatations and expansions are obtained. Apart from \eqref{109}, the other deformed generators are given by
\begin{gather}
\nonumber \nc{\cH} = \frac{\vec{\nc{p}}^2}{2m},\\
\nonumber \nc{\cP}^i = \nc{p}^i,\\
\nonumber \nc{\cG}^i = m\nc{x}^i - t\nc{p}^i + \frac{m}{2}\theta^{ij}\nc{p}^j,\\
\nonumber \nc{\cD} = \frac{1}{2}\left(\nc{p}^i\nc{x}^i + \nc{x}^i\nc{p}^i\right) - \frac{1}{m}\vec{\nc{p}}^2 t,\\
\label{112} \nc{\cK} = \frac{m}{2}\left(\nc{x}^i - \frac{\nc{p}^i}{m}t\right)^2 + \frac{m}{2}\theta^{ij}\nc{x}^i\nc{p}^j - \frac{m}{8}\theta^{i\ell}\theta^{\ell m}\nc{p}^i\nc{p}^m.
\end{gather}
These deformed generators are all consistent with \eqref{102} and satisfy the usual Schr\"odinger algebra \eqref{103}. Note that, for expansions ($\nc{\cK}$), the minimal deformation involves a term quadratic in $\theta$.

We have thus obtained the cherished deformed Schr\"odinger symmetry for a free particle on noncommutative space exactly in analogue to the Poincar\'e symmetry on noncommutative spacetime. For $\theta = 0$, the deformed generators \eqref{112} reduce to the undeformed ones given in \eqref{104}.

A more involved analysis is required for the harmonic oscillator problem which also carries a Schr\"odinger symmetry \cite{Nied-73b, Duval:1993hs}. The Hamiltonian of an $n$-dimensional oscillator is given by $\cH = \vec{p}^2/2m + m\omega^2\vec{x}^2/2$. The classical trajectories are $\vec{x}(t) = \vec{A}\cos \omega t + (1/m\omega)\vec{B}\sin \omega t$. In terms of the parameters $\vec{A}$ and $\vec{B}$,
\begin{gather}
\nonumber \vec{A} = \vec{x}\cos \omega t - \frac{\vec{p}}{m\omega}\sin\omega t,\\
\label{113} \vec{B} = m\omega\vec{x}\sin \omega t + \vec{p}\cos\omega t,
\end{gather}
the components of the moment map (generators) are given by
\begin{gather}
\nonumber \cJ^{ij} = x^i p^j - x^j p^i,\\
\nonumber \cP^i = B^i = m\omega x^i \sin \omega t + p^i \cos \omega t,\\
\nonumber \cG^i = mA^i = mx^i\cos \omega t - \frac{p^i}{\omega}\sin \omega t,\\
\nonumber \cH = \frac{1}{2m}\vec{B}^2 = \frac{1}{2}\left[m\omega^2\vec{x}^2\sin^2\omega t + \frac{\vec{p}^2}{m}\cos^2 \omega t + \frac{1}{2}\omega\left(x^i p^i + p^i x^i\right)\sin 2\omega t\right],\\
\nonumber \cM = m,\\
\nonumber \cD = \frac{1}{2}\left(B^i A^i + A^i B^i\right) = \frac{1}{2}\left(m\omega^2\vec{x}^2 - \frac{\vec{p}^2}{m\omega}\right)\sin 2\omega t + \frac{1}{2}\left(x^i p^i + p^i x^i\right)\cos 2\omega t,\\
\label{114} \cK = \frac{1}{2}m\vec{A}^2 = \frac{1}{2}m\left[\vec{x}^2\cos^2 \omega t + \frac{\vec{p}^2}{m^{2}\omega^{2}}\sin^2 \omega t - \frac{x^i p^i + p^i x^i}{2m\omega}\sin 2\omega t\right].
\end{gather}

Now we compute the deformed versions of these operators. The angular momentum is of course just given by \eqref{109}. For the other operators we observe that they appear as certain combinations of $\vec{A}$ and $\vec{B}$. Hence proper deformations of these operators should be sufficient. The canonical algebra,
\begin{equation}\label{115}
\left[\vec{A}, \vec{A}\right] = \left[\vec{B}, \vec{B}\right] = 0, \qquad \left[\vec{A}, \vec{B}\right] = \ui,
\end{equation}
is preserved by the deformations
\begin{gather}
\nonumber \nc{A}^i = \nc{x}^i\cos \omega t - \frac{\nc{p}^i}{m\omega}\sin\omega t + \frac{1}{2}\theta^{ij}\nc{p}^j \cos\omega t,\\
\label{116} \nc{B}^i = m\omega\nc{x}^i\sin \omega t + \nc{p}^i\cos\omega t + \frac{m\omega}{2}\theta^{ij}\nc{p}^j \sin\omega t.
\end{gather}
Consequently the deformed generators are given by
\begin{gather}
\nonumber \nc{\cP}^i = \nc{B}^i = m\omega \nc{x}^i \sin \omega t + \nc{p}^i \cos \omega t + \frac{1}{2}m\omega\theta^{ij}\nc{p}^j \sin \omega t,\\
\nonumber \nc{\cG}^i = m\nc{A}^i = m\nc{x}^i\cos \omega t - \frac{\nc{p}^i}{\omega}\sin \omega t + \frac{1}{2}m\theta^{ij}\nc{p}^j \cos \omega t,\\
\nonumber \begin{split}\nc{\cH} = \frac{1}{2m}\vec{\nc{B}}^2 &= \frac{1}{2}\bigg(m\omega^2\vec{\nc{x}}^2 + \frac{1}{2}m\omega^2 \theta^{ij}\left(\nc{x}^i\nc{p}^j + \nc{p}^j\nc{x}^i\right) + \frac{1}{4}m\omega^2\theta^{ij}\theta^{ik}\nc{p}^j\nc{p}^k\bigg)\sin^2\omega t\\
&\quad {} + \frac{\vec{\nc{p}}^2}{2m}\cos^2 \omega t + \frac{1}{4}\omega\left(\nc{x}^i \nc{p}^i + \nc{p}^i \nc{x}^i\right)\sin 2\omega t,\end{split}\\
\nonumber \begin{split}\nc{\cD} = \frac{1}{2}\left(\nc{B}^i \nc{A}^i + \nc{A}^i \nc{B}^i\right) &= \frac{1}{2}\bigg(m\omega^2\vec{\nc{x}}^2 - \frac{\vec{\nc{p}}^2}{m\omega} + m\omega\theta^{ij}\nc{x}^i\nc{p}^j\\
&\quad{} + \frac{m\omega}{2}\theta^{ij}\theta^{ik}\nc{p}^j\nc{p}^k\bigg)\sin 2\omega t + \frac{1}{2}\left(\nc{x}^i \nc{p}^i + \nc{p}^i \nc{x}^i\right)\cos 2\omega t,\end{split}\\
\label{117} \begin{split}\nc{\cK} = \frac{1}{2}m\vec{\nc{A}}^2 &= \frac{1}{2}m\bigg[\vec{\nc{x}}^2\cos^2 \omega t + \frac{\vec{\nc{p}}^2}{m^{2}\omega^{2}}\sin^2 \omega t - \frac{\nc{x}^i \nc{p}^i + \nc{p}^i \nc{x}^i}{2m\omega}\sin 2\omega t\\
&\qquad\qquad {} + \theta^{ij}\nc{x}^i \nc{p}^j\cos^2 \omega t + \frac{1}{4}\theta^{ij}\theta^{ik}\nc{p}^j\nc{p}^k\cos^2 \omega t\bigg].\end{split}
\end{gather}
All these are consistent with \eqref{102} and satisfy the Schr\"odinger algebra \eqref{103}. The basic commutators \eqref{102} and \eqref{106} are needed to check this algebra.

In the $\omega \rightarrow 0$ limit the harmonic oscillator goes over to the free particle. It is easy to verify that, in this limit, the deformed generators \eqref{117} reduce to those found in \eqref{112} for the free-particle case.


\subsection*{2. Applications}

Here we consider some applications of the general formalism.

\subsubsection*{Deformed generators in different representations}

In the usual commutative space a symmetry exists between the coordinates $x$ and momenta $p$. Each is an observable with eigenvalues extending from $-\infty$ to $+\infty$ and the usual commutation relations involving $x$ and $p$ remain invariant if $x$ and $p$ are interchanged and `$\ui$' is replaced by `$-\ui$'. One may then set up the coordinate representation in which $x$ is diagonal and $p =-\ui \frac{\partial}{\partial x}$ with $\hbar=1$. Alternatively it is also feasible to write the momentum representation where $p$ is diagonal and $x = \ui \frac{\partial}{\partial p}$.

In the noncommutative space, on the other hand, the symmetry between $x$ and $p$ is lost and some nontrivial differences among the two representations are expected. From a purely algebraic point of view one may use either representation. However if a wavefunction is considered that can be expanded in a complete commuting set of observables (the basis vectors), then the momentum representation is singled out. It appears that, for noncommutative space, momentum representation is more favoured. This is even true from an algebraic point of view, as we now demonstrate. For simplicity we confine to the Galilean sector.

The Galilean generators \eqref{109}, \eqref{112}, in the coordinate representation, $\nc{x}^i = \nc{x}^i$, $\nc{p}^i = -\ui \frac{\partial}{\partial \nc{x}^i}$, are given by
\begin{gather}
\nonumber \nc{\cJ}^{ij} = -\ui \nc{x}^i \frac{\partial}{\partial \nc{x}^j} + \ui  \nc{x}^j\frac{\partial}{\partial \nc{x}^i} - \frac{1}{2}\theta^{im}\frac{\partial}{\partial \nc{x}^m} \frac{\partial}{\partial \nc{x}^j} + \frac{1}{2}\theta^{jm}\frac{\partial}{\partial \nc{x}^m} \frac{\partial}{\partial \nc{x}^i},
\\
\nonumber \nc{\cH} = - \frac{1}{2m}\frac{\partial^2}{{\partial \nc{x}}^2},
\\
\nonumber \nc{\cP}^i = -\ui \frac{\partial}{\partial \nc{x}^i},
\\
\label{YY-2} \nc{\cG}^i = m\nc{x}^i + \ui t \frac{\partial}{\partial \nc{x}^i} -\ui  \frac{m}{2}\theta^{ij} \frac{\partial}{\partial \nc{x}^j},
\end{gather}
where the derivatives satisfy the algebra \cite{Wess-04}
\begin{equation} \label{YY-1}
\left[\frac{\partial}{\partial \nc{x}^i}, \frac{\partial}{\partial \nc{x}^j}\right] = 0, \qquad
\left[\frac{\partial}{\partial \nc{x}^i}, \nc{x}^j\right] = \delta^{ij}.
\end{equation}
Using these relations and taking into account \eqref{102}, the generators \eqref{YY-2} are seen to satisfy the usual Galilean algebra.

In the momentum representation, on the other hand, we have
\[
\nc{p}^i = \nc{p}^i, \qquad \nc{x}^i = \ui \frac{\partial}{\partial \nc{p}^i} - \frac{1}{2} \theta^{ij} \nc{p}^j,
\]
where the derivative satisfies commutation rules similar to \eqref{YY-1}:
\[
\left[ \frac{\partial}{\partial \nc{p}^i}, \frac{\partial}{\partial \nc{p}^j} \right] = 0, \qquad
\left[ \frac{\partial}{\partial \nc{p}^i}, \nc{p}^j \right] = \delta^{ij}.
\]
Note that there is a deformation in the representation of $\nc{x}^i$ to correctly reproduce \eqref{102} and \eqref{106}. It also shows the loss of symmetry between two representations.

The deformed rotation operator \eqref{109} now has the form
\begin{equation*}\label{109m}
\begin{split}
\nc{\cJ}^{ij} &= \nc{p}^j \nc{x}^i - \nc{p}^i \nc{x}^j + \frac{1}{2}\theta^{im}\nc{p}^m\nc{p}^j - \frac{1}{2}\theta^{jm}\nc{p}^m\nc{p}^i
\\
&= \ui \nc{p}^j \frac{\partial}{\partial \nc{p}^i} - \ui \nc{p}^i \frac{\partial}{\partial \nc{p}^j}.
\end{split}
\end{equation*}
The $\theta$-dependent terms cancel out completely. The same happens for the boost and we get
\[
\nc{\cG}^i =  \ui m \frac{\partial}{\partial \nc{p}^i} - t\nc{p}^i.
\]
We thus find that all the generators have exactly the same structure as in the commutative description.\footnote{This is also true for the conformal generators. Even for the Poincar\'e case the same feature persists as we have checked by considering the expressions given in \cite{Wess-04, Dimitrijevic:2004rf}.} The validity of the Galilean (or the Schr\"odinger) algebra, in this representation becomes obvious. It shows the naturalness of the momentum representation---a fact which we had argued on more general considerations.

\subsubsection*{Inonu--Wigner contraction and second central extension}

It has been known for a long time that the Galilei group in $(2+1)$ dimensions admits a two-parameter central extension \cite{Levy-72:Grig-96:Bose-95}. One of these is the mass $m$ which is present in all dimensions. The other is somewhat exotic and confined specifically to two space dimensions. It corresponds to spin which can be any real number \cite{Duva-00, Jack-00, Horvathy:2002vt}.

Here we show that deformed Galilean generators allow a general type of algebra leading to a second central extension that persists in any (and not just two) dimensions. Furthermore such an algebra is also obtained by means of an Inonu--Wigner contraction of the Poincar\'e algebra.

Let us consider the general form for boosts given in \eqref{111} with $\lambda_2 = 0$ and $\lambda_1 = \lambda$:
\begin{equation} \label{111-gen}
\ncg{\cG}^i = m\nc{x}^i - t \nc{p}^i + \lambda m \theta^{ij} \nc{p}^j.
\end{equation}
Correspondingly, a generalised version of the rotation operator \eqref{109} is written as
\begin{equation} \label{109-gen}
\ncg{\cJ}^{ij} = \nc{x}^i \nc{p}^j - \nc{x}^j \nc{p}^i + \lambda \left( \theta^{im} \nc{p}^j - \theta^{jm} \nc{p}^i \right) \nc{p}^m.
\end{equation}
Then the algebra of these operators is given by
\begin{gather} \label{121-a}
\left[ \ncg{\cG}^i, \ncg{\cG}^j \right] = (1 - 2\lambda) \ui m^2 \theta^{ij},
\\
\label{xxxx-q}
\left[ \ncg{\cG}^{i}, \ncg{\cJ}^{k\ell} \right]
= \ui \left( \delta^{i\ell} \ncg{\cG}^{k} - \delta^{ik} \ncg{\cG}^{\ell} \right)
+ \ui m (1 - 2 \lambda) \left( \theta^{ik} \nc{p}^{\ell} - \theta^{i\ell} \nc{p}^{k} \right),
\\
\label{xxxx-qA}
\begin{split}
\left[ \ncg{\cJ}^{ij}, \ncg{\cJ}^{k\ell} \right]
&= \ui \left( \delta^{ik} \ncg{\cJ}^{j\ell} - \delta^{jk} \ncg{\cJ}^{i\ell} + \delta^{j\ell} \ncg{\cJ}^{ik} - \delta^{i\ell} \ncg{\cJ}^{jk} \right)
\\
&\quad {} + \ui (1 - 2 \lambda) \left( \theta^{ik} \nc{p}^{j} \nc{p}^{\ell} -  \theta^{jk} \nc{p}^{i} \nc{p}^{\ell} - \theta^{i\ell} \nc{p}^{j} \nc{p}^{k} + \theta^{j\ell} \nc{p}^{i} \nc{p}^{k} \right).
\end{split}
\end{gather}
The algebra with the momentum and the Hamiltonian is unchanged. We find that the boosts do not commute which yields the second central extension. This is true in any dimensions and not just for two where it is usually observed \cite{Levy-72:Grig-96:Bose-95, Duva-00, Jack-00, Horvathy:2002vt}.

It is possible to show that the above generalised algebra is obtained from a group contraction of a generalised Poincar\'e algebra recently discussed in \cite{Koch-05}. There the angular momentum is defined as
\begin{equation} \label{109-Koch}
\nc{\cJ}^{\mu \nu} = \nc{x}^\mu \nc{p}^\nu - \nc{x}^\nu \nc{p}^\mu + \lambda \left( \theta^{\mu \alpha} \nc{p}^\nu - \theta^{\nu \alpha} \nc{p}^\mu \right) \nc{p}^\alpha,
\end{equation}
which yields \cite{Koch-05}
\begin{equation} \label{xxxx-qAKoch}
\begin{split}
\left[ \ncg{\cJ}^{\mu \nu}, \ncg{\cJ}^{\rho \sigma} \right]
&= \ui \left( \delta^{\mu \rho} \ncg{\cJ}^{\nu \sigma} - \delta^{\nu \rho} \ncg{\cJ}^{\mu \sigma} + \delta^{\nu \sigma} \ncg{\cJ}^{\mu \rho} - \delta^{\mu \sigma} \ncg{\cJ}^{\nu \rho} \right)
\\
&\quad {} + \ui (1 - 2 \lambda) \left( \theta^{\mu \rho} \nc{p}^{\nu} \nc{p}^{\sigma} -  \theta^{\nu \rho} \nc{p}^{\mu} \nc{p}^{\sigma} - \theta^{\mu \sigma} \nc{p}^{\nu} \nc{p}^{\rho} + \theta^{\nu \sigma} \nc{p}^{\mu} \nc{p}^{\rho} \right).
\end{split}
\end{equation}
In performing the group contraction from Poincar\'e to Galileo, observe that for a system of particles of mass $m$ and velocity $v$, the various operators are expected to be of the following orders \cite{Wein}: $\nc{\cJ}^i \sim 1$, $\nc{\cP}^i \sim mv$, $\nc{\cJ}^{0i} = - \nc{\cG}^i \sim 1/v$, $\theta^{ij} \sim 1 / m^2 v^2$, $\theta^{0i} \sim 0$, $\cP^0 = \cH \sim m + mv^2$ (energy $\sim$ mass energy + kinetic/potential energy) and finally the limit $v \ll 1$ has to be taken. The space component of \eqref{xxxx-qAKoch} contract to \eqref{xxxx-qA}. Likewise the other relations \eqref{121-a} and \eqref{xxxx-q} also follow on taking appropriate limits.

It is possible to discuss another generalisation of the Galilean algebra such that, for two dimensions, the usual algebra (but with noncommuting boosts) is obtained. This happens if the boost and angular momentum are taken as \eqref{111-gen} and \eqref{109}.  Then the only modifications are in the algebra of boosts (once again given by \eqref{121-a}) and \eqref{xxxx-q} is replaced by
\begin{equation} \label{xxxx}
\left[ \nc{\cG}^{i}, \nc{\cJ}^{k\ell} \right]
= \ui \left( \delta^{i\ell}\nc{\cG}^{k} - \delta^{ik}\nc{\cG}^{\ell} \right)
+ \ui m \left( \lambda - \frac{1}{2} \right) \left( \delta^{ik} \theta^{\ell j} \nc{p}^{j} - \delta^{i\ell} \theta^{kj} \nc{p}^{j} - \theta^{ik} \nc{p}^{\ell} - \theta^{i\ell} \nc{p}^{k} \right).
\end{equation}
The other brackets conform to the usual Galilean algebra. Note that the algebra closes although it is distinct from the standard Galilean algebra. For two dimensions, however, the additional piece vanishes. We get back the usual Galilean algebra but the second central extension found in the algebra of boosts persists.

To conclude, the deformed Schr\"odinger operators on noncommutative space were obtained. These operators preserved the standard Schr\"odinger algebra. For vanishing $\theta^{ij}$, the undeformed operators were reproduced from the deformed ones. The structures of the generators in the coordinate and momentum representations were derived. In the latter case there was a form invariance; i.e., the generators had the same form as in the usual (commutative) description. Hence computations in theories in noncommutative space are considerably simplified in the momentum representation. It was possible to discuss a generalised Galilean algebra that was also obtained from a contraction of a generalised Poincar\'e algebra. The commutator of the boosts was nonvanishing leading to a second central extension valid in any dimensions. We discussed another type of generalised Galilean algebra which, in two dimensions, reduced to the usual form but retained the central extension in the commutator of boosts. Thus our deformed generators provided an alternative way of understanding the second central extension of the planar Galilei group.







\begin{thebibliography}{99}\setlength{\itemsep}{0.0cm}\small \raggedright
\bibitem{Wess-04} J.~Wess, ``Deformed coordinate spaces: Derivatives,'' hep-th/0408080.
\bibitem{Dimitrijevic:2004rf} M.~Dimitrijevic and J.~Wess, ``Deformed bialgebra of diffeomorphisms,'' hep-th/0411224.
\bibitem{Asch-05} P.~Aschieri, C.~Blohmann, M.~Dimitrijevic, F.~Meyer, P.~Schupp and J.~Wess, 
 Class.\ Quant.\ Grav.\ 22 (2005) 3511 [hep-th/0504183].
\bibitem{Koch-05} F.~Koch and E.~Tsouchnika, 
 Nucl.\ Phys.\ B 717 (2005) 387 [hep-th/0409012].
\bibitem{Chai-05} M.~Chaichian, P.~P.~Kulish, K.~Nishijima and A.~Tureanu, 
 Phys.\ Lett.\ B 604 (2004) 98 [hep-th/0408069]; M.~Chaichian, P.~Pre\v{s}najder and A.~Tureanu, 
 Phys.\ Rev.\ Lett.\  94 (2005) 151602 [hep-th/0409096].
\bibitem{Matlock:2005zn} P.~Matlock, 
 Phys.\ Rev.\ D 71 (2005) 126007 [hep-th/0504084].
\bibitem{Oeckl:2000eg} R.~Oeckl, 
 Nucl.\ Phys.\ B 581 (2000) 559 [hep-th/0003018].
\bibitem{Banerjee:2004ev} R.~Banerjee, B.~Chakraborty and K.~Kumar, 
 Phys.\ Rev.\ D 70 (2004) 125004 [hep-th/0408197].
\bibitem{Gonera:2005hg} C.~Gonera, P.~Kosinski, P.~Maslanka and S.~Giller, 
 Phys.\ Lett.\ B 622 (2005) 192 [hep-th/0504132].
\bibitem{Luki-05} J.~Lukierski and M.~Woronowicz, ``New Lie-algebraic and quadratic deformations of Minkowski space from twisted Poincare symmetries,'' hep-th/0508083.
\bibitem{Nied-73:Hage-72:Burd-72} U.~Niederer, Helv. Phys. Acta 45 (1972) 802; C.~R.~Hagon, Phys. Rev. D 5 (1972) 377; G.~Burdet and M.~Perrin, Lett. Nuovo Cimento 4 (1972) 651.
\bibitem{Levy-72:Grig-96:Bose-95} J.~M.~L\'evy-Leblond in E.~Loebl, Group Theory and Applications, Academic Press, New York (1972); D.~R.~Grigore, J. Math. Phys. 37 (1996) 460; S.~K.~Bose, Comm. Math. Phys. 169 (1995) 385.
\bibitem{Duva-00} C.~Duval and P.~A.~Horvathy, 
 Phys.\ Lett.\ B 479 (2000) 284 [hep-th/0002233].
\bibitem{Nied-73b} U.~Niederer, Helv. Phys. Acta 46 (1973) 192.
\bibitem{Duval:1993hs} C.~Duval and P.~A.~Horvathy, 
 J.\ Math.\ Phys.\  35 (1994) 2516 [hep-th/0508079].
\bibitem{Jack-00} R.~Jackiw and V.~P.~Nair, 
 Phys.\ Lett.\ B 480 (2000) 237 [hep-th/0003130].
\bibitem{Horvathy:2002vt} P.~A.~Horvathy and M.~S.~Plyushchay, 
 JHEP 06 (2002) 033 [hep-th/0201228].
\bibitem{Wein} S. Weinberg, ``The Quantum Theory of Fields,'' Vol.~1, \textit{Cambridge University Press}, Cambridge, 1995.
\end{thebibliography}
\end{document}